\documentclass[11pt,a4paper]{article}
\pdfoutput=1
\usepackage{jheppub}

\usepackage{multirow, graphicx,amssymb,url,mathrsfs,amsmath}
\usepackage{wrapfig,boxedminipage,setspace,subfigure,epsfig}
\usepackage{amsxtra,amstext,latexsym,dsfont,amsfonts}
\usepackage{color,eucal}
\usepackage[dvipsnames]{xcolor}
\usepackage{float}
\usepackage{slashed,comment}
\usepackage{kotex}

\usepackage{tikz}
\usetikzlibrary{calc,patterns,angles,quotes}
\usetikzlibrary{decorations.pathreplacing,decorations.markings,snakes}

\usepackage{tabularx, array}
\newcolumntype{L}[1]{>{\raggedright\arraybackslash}p{#1}}
\newcolumntype{C}[1]{>{\centering\arraybackslash}p{#1}}
\newcolumntype{R}[1]{>{\raggedleft\arraybackslash}p{#1}}

\newcommand{\dd}{\mathrm{d}}

\newcommand{\bra}[1]{\mbox{$\langle #1 |$}}
\newcommand{\ket}[1]{\mbox{$| #1 \rangle$}}

\newcommand{\braket}[2]{\mbox{$\langle #1  | #2 \rangle$}}

\title{Spread complexity in saddle-dominated scrambling}

\author[a]{Kyoung-Bum Huh,}
\author[b,c]{Hyun-Sik Jeong}
\author[b]{and Juan F. Pedraza}

\emailAdd{hkabell1689@gm.gist.ac.kr}
\emailAdd{hyunsik.jeong@csic.es}
\emailAdd{j.pedraza@csic.es}

\preprint{IFT-UAM/CSIC-23-178}

\affiliation[a]{Department of Physics and Photon Science, Gwangju Institute of Science and Technology,\\123 Cheomdan-gwagiro, Gwangju 61005, Korea}
\affiliation[b]{Instituto de F\'isica Te\'orica UAM/CSIC, Calle Nicol\'as Cabrera 13-15, 28049 Madrid, Spain}
\affiliation[c]{Departamento de F\'isica Te\'orica, Universidad Aut{\'o}noma de Madrid, 
28049 Madrid, Spain}

\abstract{
Recently, the concept of spread complexity, Krylov complexity for states, has been introduced as a measure of the complexity and chaoticity of quantum systems. In this paper, we study the spread complexity of the thermofield double state within \emph{integrable} systems that exhibit saddle-dominated scrambling. Specifically, we focus on the Lipkin-Meshkov-Glick model and the inverted harmonic oscillator as representative examples of quantum mechanical systems featuring saddle-dominated scrambling. Applying the Lanczos algorithm, our numerical investigation reveals that the spread complexity in these systems exhibits features reminiscent of \emph{chaotic} systems, displaying a distinctive ramp-peak-slope-plateau pattern. Our results indicate that, although spread complexity serves as a valuable probe, accurately diagnosing true quantum chaos generally necessitates additional physical input. We also explore the relationship between spread complexity, the spectral form factor, and the transition probability within the Krylov space. We provide analytical confirmation of our numerical results, validating the Ehrenfest theorem of complexity and identifying a distinct quadratic behavior in the early-time regime of spread complexity.
}

\begin{document}
\maketitle

\section{Introduction}\label{}
Classical chaos is a widespread phenomenon in nature, describing the unpredictable behavior of deterministic dynamical systems due to their exponential sensitivity to initial conditions. This concept has foundational implications in physics, notably in the fields of thermodynamics and hydrodynamics~\cite{Ott,Gaspard}.  In contrast, \textit{quantum chaos} currently lacks a precise definition, encompassing various interpretations stemming from numerous endeavors to extend the notion of chaos into the quantum realm.

Historically, defining quantum chaos has been challenging, leading to various attempts using different probes and measures. Currently, the most versatile and widely used approach involves examining the level statistics of the quantum mechanical Hamiltonian~\cite{Casati1980,Bohigas:1984aa,10.2307/79349,BERRY1981163,Santos:2010aa,DAlessio:2016aa}. Chaotic systems typically display statistical patterns consistent with Wigner-Dyson statistics, while integrable systems show characteristics reminiscent of a Poisson distribution.

While level statistics effectively diagnose the appearance of chaos over sufficiently long time scales, understanding its early development demands additional probes that are sensitive to shorter time scales. An indirect yet effective method to access this regime involves scrutinizing \textit{operator growth}. In chaotic systems, it is expected that operators exhibit a more rapid growth compared to their integrable counterparts. This approach offers a convenient means to explore and characterize the dynamics of chaos on shorter time scales, contributing valuable insights to the broader understanding of quantum chaotic systems.

\paragraph{Out-of-time-order correlators.}
In the pursuit of quantifying operator growth, considerable attention has been dedicated to exploring the role of out-of-time-order correlators (OTOCs)~\cite{Larkin1969QuasiclassicalMI} as promising indicators for detecting the early development of quantum chaos. Indeed, OTOCs are envisioned as valuable probes within chaotic systems, potentially exhibiting exponential growth over time —a characteristic attributed as a signature marking the onset of quantum chaos. This parallels the classical notion of chaos, providing an insightful analogy for understanding the evolving dynamics in quantum systems.

The rate of such exponential growth is quantified by the so-called (quantum) Lyapunov exponent $\lambda_{L}$, which has an upper limit set by the Maldacena-Shenker-Stanford (MSS) bound~\cite{Maldacena2016,Murthy:2019fgs,Avdoshkin:2019trj}: $\lambda_{L}\leq 2\pi T$, where $T$ denotes temperature. Systems exhibiting maximal chaos, i.e., those saturating the bound, have been specifically highlighted as illustrative models for understanding strongly coupled systems with Einstein gravity duals~\cite{Sachdev:1992fk,Maldacena:2016hyu,Sachdev:2015efa,Kitaev:2017awl}.\footnote{It is also noteworthy that the proposed definition of quantum chaos in terms of OTOCs, has been driven by a broader research program encompassing quantum information, black holes, and holography~\cite{Hayden_2007,Sekino:2008he,Lashkari:2011yi,Shenker:2013pqa,Shenker2015,Yoshida:2017non,deBoer:2017xdk,Yoshida:2018vly,Landsman:2018jpm}.}

However, recent inquiries have cast doubt on the significance of the OTOC. Specifically, the exponential growth may not unequivocally indicate the chaotic nature of the system, as highlighted in various studies~\cite{Pilatowsky-Cameo:2019qxt,Xu:2019lhc,Rozenbaum:2019nwn,Hashimoto:2017oit,Hashimoto:2020xfr,Bhattacharyya:2020art}. This phenomenon can occur simply as a result of unstable saddle points in the classical phase space~\cite{Xu:2019lhc}. Thus, it is crucial to differentiate this phenomenon, termed \emph{saddle-dominated scrambling}, from genuine quantum chaos.

In essence, the exponential growth of OTOCs can manifest even in integrable systems. Thus, relying solely on the exponential growth of OTOCs proves insufficient as a standalone indicator for identifying quantum chaos. Notably, well-established toy models exemplifying this phenomenon (saddle-dominated scrambling) include the Lipkin-Meshkov-Glick (LMG) model~\cite{Xu:2019lhc} and the inverted harmonic oscillator~\cite{Hashimoto:2020xfr}.\footnote{See also \cite{Bhattacharyya:2020art,Qu:2021ius,Qu:2022zwq} for the circuit complexity and \cite{Trunin:2023xmw,Trunin:2023rwm} for the redefined OTOCs.}

\paragraph{Krylov (operator/state) complexity.}
In recent years, a promising alternative probe for the direct and quantitative assessment of operator growth, known as Krylov ``operator'' complexity, has emerged~\cite{Parker:2018yvk}. The fundamental concept involves generating the Krylov basis and Lanczos coefficients $b_n$ using the Lanczos algorithm. This approach enables the definition of Krylov operator complexity, a direct measure of the rate of growth of initial operators under Heisenberg time evolution.\footnote{Note that, in computing the Krylov complexity, a given quantum system would be reduced to a one-dimensional Krylov chain model, wherein the hopping is determined by the Lanczos coefficients.}

It has been suggested that Lanczos coefficients experience their most rapid growth in chaotic systems \cite{Parker:2018yvk}. In chaotic systems, the Lanczos coefficients follow a linear growth ($\approx \alpha n$), resulting in an exponential growth in Krylov operator complexity ($\approx e^{2 \alpha t}$). An intriguing connection between OTOCs and Krylov operator complexity arises, as the growth rate of the Lanczos coefficient serves as an upper bound for the Lyapunov exponent \cite{Parker:2018yvk}: $\lambda_L \leq 2\alpha$. Consequently, the growth of Lanczos coefficients may provide a more stringent bound than the Maldacena-Shenker-Stanford (MSS) bound \cite{Avdoshkin:2022xuw}: $\lambda_L \leq 2\alpha \leq 2\pi T$.

More recently, a further development in the realm of complexity measures has emerged. This development, termed Krylov ``state'' complexity, or \textit{spread complexity}, builds upon the established concept of Krylov operator complexity and serves as a quantitative measure for assessing the complexity of a quantum state in the Schrödinger picture~\cite{Balasubramanian:2022tpr}. 
Spread complexity is defined to quantify the spread of a wave function minimized across all possible bases within the Hilbert space. Importantly, it has been demonstrated \cite{Balasubramanian:2022tpr} that such a minimum can be uniquely obtained by the Krylov basis.

It is noteworthy to recognize that both Krylov operator complexity and spread complexity capture their time evolution by extracting information from the Lanczos coefficients. However, two key distinctions emerge. Firstly, unlike Krylov operator complexity, spread complexity involves two types of Lanczos coefficients ($a_n, b_n$). However, their precise role in physics remains somewhat unclear. That is, these Lanczos coefficients do not seem to have a clear association with the Lyapunov exponent. Secondly, in contrast to Krylov operator complexity, spread complexity does not demonstrate exponential growth in chaotic systems. Instead, a different conjecture is proposed for chaotic systems~\cite{Balasubramanian:2022tpr}. Regarding the time-evolved thermofield double states in chaotic systems, spread complexity exhibits four distinct regimes: a linear ramp leading to a \textit{peak}, followed by a downward slope to a plateau. In integrable systems, the peak diminishes.

A few comments regarding the aforementioned conjecture are in order. Firstly, the linear ramp, peak, slope, and plateau observed in spread complexity bear an analogy to the slope, dip, ramp, and plateau seen in the spectral form factor (for example, refer to Fig. \ref{LMGComvsSFF}). This analogy is further explained in \cite{Balasubramanian:2022tpr,Erdmenger:2023wjg}.
Secondly, the initial linear ramp and plateau align with the anticipated behavior for complexity in chaotic systems, as conjectured in \cite{Susskind:2014rva}. The authors of \cite{Balasubramanian:2022tpr} propose that the peak followed by a downward slope could also be universal features of complexity dynamics in chaotic systems.
Lastly, in the case of a maximally entangled state, such as the thermofield double state, spread complexity relies solely on the Hamiltonian's spectrum. It has been demonstrated that, at late times, a non-trivial relationship exists between spread complexity and the spectral form factor~\cite{Balasubramanian:2022tpr,Erdmenger:2023wjg}.

Beyond its potential as a tool for probing quantum chaos, Krylov complexity has spurred extensive research due to its non-trivial connection with the spectral form factor and its implications for holography.\footnote{The concept of Krylov complexity is gaining prominence, particularly in the context of holography, as it is well-defined in any quantum theory, overcoming the ambiguities in defining specific elementary gates or establishing tolerances.} This interest has led to investigations across various domains, including research in diverse quantum systems~\cite{Balasubramanian:2022tpr,Barbon:2019wsy,Avdoshkin:2019trj,Dymarsky:2019elm,Rabinovici:2020ryf,Cao:2020zls,Kim:2021okd,Rabinovici:2021qqt,Trigueros:2021rwj,Fan:2022xaa,Heveling:2022hth,Bhattacharjee:2022vlt,Caputa:2022eye,Caputa:2022yju,Muck:2022xfc,Rabinovici:2022beu,He:2022ryk,Hornedal:2022pkc,Alishahiha:2022anw,Baek:2022pkt,Bhattacharjee:2023dik,Bhattacharyya:2023dhp,Erdmenger:2023wjg,Hashimoto:2023swv,Camargo:2023eev,Balasubramanian:2023kwd,Bhattacharya:2023yec,Bhattacharjee:2022qjw,Caputa:2024vrn}, gauge theories~\cite{Magan:2020iac}, holographic models~\cite{Jian:2020qpp,Rabinovici:2023yex}, conformal/quantum field theories~\cite{Dymarsky:2021bjq,Caputa:2021ori,Avdoshkin:2022xuw,Camargo:2022rnt,Adhikari:2022whf,Vasli:2023syq}, Lie groups~\cite{Caputa:2021sib,Patramanis:2021lkx,Patramanis:2023cwz,Chattopadhyay:2023fob}, matrix models~\cite{Iizuka:2023pov,Iizuka:2023fba}, models of quantum quenches~\cite{Afrasiar:2022efk,Pal:2023yik} and open quantum systems~\cite{Bhattacharya:2022gbz,Liu:2022god,Bhattacharjee:2022lzy,Bhattacharya:2023zqt,Bhattacharjee:2023uwx}.

\paragraph{Motivation of the paper.}
In this paper, we investigate the Krylov complexity within quantum mechanical models, with a specific focus on the analysis of spread complexity. Our investigation centers on integrable models that showcase unstable saddle points, a phenomenon recognized as saddle-dominated scrambling. To illustrate this, we consider the Lipkin-Meshkov-Glick (LMG) model and the inverted harmonic oscillator, drawing parallels with previous analyses of OTOCs found in~\cite{Xu:2019lhc, Hashimoto:2020xfr}.

It is crucial to highlight a parallel investigation of Krylov operator complexity in these models, presented in \cite{Bhattacharjee:2022vlt, Baek:2022pkt}. In these papers, it was shown that, despite the integrability of the models, Krylov operator complexity can exhibit exponential growth. Consequently, both the OTOCs and Krylov operator complexity face challenges in distinguishing true quantum chaos, as they yield similar imprints in integrable models featuring saddle-dominated scrambling.  This raises the question of whether spread complexity holds a more discerning capability in differentiating between saddle-dominated scrambling and true quantum chaos, serving as our primary motivation. To the best of our knowledge, this aspect has not been previously acknowledged or thoroughly investigated.

{
Furthermore, it is noteworthy that a recent study \cite{Caputa:2024vrn} provides a thorough examination of the comparison and contrast between Krylov operator complexity and spread complexity. Their findings reveal that in general both complexities can exhibit disparate behaviors, particularly in the intermediate time regime where the peak of spread complexity becomes manifest. Our work thus can stand as a distinct and original contribution to the field, separate from the analyses of Krylov operator complexity conducted in \cite{Bhattacharjee:2022vlt, Baek:2022pkt}.
}

In addition to validating the conjectured proposal of spread complexity for quantum chaos, which encompasses the anticipated linear-peak-slope-plateau features, we conduct a comprehensive examination of various aspects of spread complexity, including a comparative analysis between spread complexity and the spectral form factor. Moreover, we investigate another universal behavior of spread complexity concerning Hamiltonians describing chaotic systems, proposed originally in \cite{Erdmenger:2023wjg}. Specifically, we show that the transition probability, determined by the wave function in the Krylov space, sheds light on the presence (or absence) of the peak in spread complexity for chaotic (or integrable) systems.

\paragraph{Structure of the paper.}
This paper is organized as follows. 
In Sec. \ref{sec2}, we present preliminary details regarding the spread complexity of a given Hamiltonian. This includes a comprehensive review, encompassing the formalism of the Lanczos algorithm and the Ehrenfest theorem of spread complexity. Additionally, we derive an analytical expression for spread complexity at early times, presenting a novel contribution to the existing literature. Then, using the Lanczos algorithm, we present an illustrative quantum mechanical example featuring billiard systems. This example demonstrates how spread complexity can exhibit a peak in chaotic systems while remaining absent in integrable systems.
In Sec. \ref{sec3}, we introduce the Lipkin-Meshkov-Glick (LMG) model and revisit its unstable saddle points in phase space. We then compute the spread complexity of the LMG model and assess its effectiveness as a probe of quantum chaos. Additionally, we discuss its relationship with the spectral form factor and the transition probability.
In Sec. \ref{sec4}, we conduct an analysis of spread complexity within the inverted harmonic oscillator, drawing comparisons with the LMG model analyzed in Sec. \ref{sec3}. Finally, in Section \ref{sec5} we present our main conclusions and provide some interesting directions for future work.

\section{Preliminaries}\label{sec2}

\subsection{Spread complexity and Lanczos algorithm: a quick review}\label{SCLA}
In this section, we review the spread complexity~\cite{Balasubramanian:2022tpr}.
In a quantum system with Hamiltonian $H$, the Schrödinger state
\begin{align}\label{SRS}
\begin{split}
\ket{\psi(t)} = e^{-i H t} \ket{\psi(0)} \,,
\end{split}
\end{align}
can be expressed as a linear combination of $\{\ket{\psi(0)}\,, H \ket{\psi(0)}\,, H^2 \ket{\psi(0)}\,, \cdots\}$.

\paragraph{Krylov basis and Lanczos algorithm.}
Using the Gram-Schmidt procedure together with the natural inner product, we can orthonormalize \eqref{SRS} by the Lanczos algorithm:
\begin{align}\label{LA}
\begin{split}
\ket{A_{n+1}} = \left(H-a_{n}\right)\ket{K_{n}} - b_{n}\ket{K_{n-1}} \,, \qquad \ket{K_n} = b_{n}^{-1} \ket{A_n} \,,
\end{split}
\end{align}
with the initial condition 
\begin{align}\label{IC}
\begin{split}
b_0:=0 \,, \qquad \ket{K_0}:=\ket{\psi(0)}  \,.
\end{split}
\end{align}
Here, the Lanczos coefficients $a_n$ and $b_n$ are defined as
\begin{align}\label{LCE}
\begin{split}
a_n = \bra{K_n}H\ket{K_n} \,, \qquad b_n = \braket{A_n}{A_n}^{1/2}\,,
\end{split}
\end{align}

The generated orthonormal basis $\{ \ket{K_n} \}$ is called the Krylov basis where it may expand the full Hilbert space for a chaotic model~\cite{Balasubramanian:2022tpr}. It is worth noting that the Lanczos algorithm \eqref{LA} can be expressed as 
\begin{align}\label{LA2}
\begin{split}
H\ket{K_{n}} = a_{n}\ket{K_{n}} + b_{n+1} \ket{K_{n+1}} + b_{n}\ket{K_{n-1}} \,, 
\end{split}
\end{align}
which implies that the Hamiltonian can be a tridiagonal matrix in the Krylov basis, which this is also known as the Hessenberg form of the Hamiltonian. In other words, once the Hessenberg form is given, one can easily find the Lanczos coefficients.\footnote{The Hessenberg form can be computed using Householder reflections in $\textit{Mathematica}$.}

Furthermore, as proposed in \cite{Hashimoto:2023swv, Parlett_1998}, one can also modify the Lanczos algorithm \eqref{LA} in order to reduce the numerical errors in the orthogonalization as
\begin{enumerate}
\item{Define $\mathcal{D} := \text{diag}\left(E_1\,, \cdots\,, E_{N_{max}}\right)$, where $E_n$ is the eigenvalue of the given Hamiltonian $H$ and $N_{max}$ a chosen truncation number.}
\item{$b_0:=0$\,, \quad $\ket{K_0} := \ket{\psi(0)}$\,, \quad $a_0 := \bra{K_0} \mathcal{D} \ket{K_0}$.}
\item{For $n\geq1$: $\ket{A_n} = \left( \mathcal{D}-a_{n-1} \right)\ket{K_{n-1}}-b_{n-1} \ket{K_{n-2}}$.}
\item{Replace as $\ket{A_n} \rightarrow \ket{A_n} - \sum_{m=0}^{n-1} \, \braket{A_m}{K_0} \, \ket{A_m}$.}
\item{Set $b_n = \braket{A_n}{A_n}^{1/2}$.}
\item{If $b_n=0$ stop; otherwise set $\ket{K_n} = b_n^{-1} \ket{A_n} \,, a_n = \bra{K_n} \mathcal{D} \ket{K_n}$, and go to step 3.}
\end{enumerate}
In all examples examined within this paper, we verified that the outcomes (Lanczos coefficients and complexity) derived from the Hessenberg form align consistently with those obtained through this modified algorithm.

\paragraph{Krylov basis and Spread complexity.}
Using the Krylov basis $\ket{K_n}$ obtained, one can expand the Schrödinger state $\ket{\psi(t)}$ as
\begin{align}\label{SSS}
\begin{split}
\ket{\psi(t)} = \sum_{n=0} \psi_n(t) \ket{K_n} \,, 
\end{split}
\end{align}
which produces the following the Schrödinger equation
\begin{align}\label{SEQ}
\begin{split}
i \, \partial_t \psi_n(t)  = a_n \psi_n(t) + b_{n+1} \psi_{n+1}(t) + b_{n} \psi_{n-1}(t)  \,, 
\end{split}
\end{align}
where we have the initial condition $\psi_{n}(0) = \delta_{n0}$ by definition. Then, the spread complexity of the state $\ket{\psi(0)}$ is defined as 
\begin{align}\label{SCDEF}
\begin{split}
C(t) := \sum_{n=0} \, n |\psi_n(t)|^2  \,,  \qquad \sum_{n=0} \,|\psi_n(t)|^2 = 1 \,,
\end{split}
\end{align}
where it measures the average depth of a time evolving state in the Krylov basis, i.e., the spread of the wave function in the Krylov basis~\cite{Balasubramanian:2022tpr}.

\paragraph{Spread complexity of thermofield double state.}
In this paper, we investigate the growth of spread complexity for Thermofield double (TFD) state as in the original literature~\cite{Balasubramanian:2022tpr}, which is defined as 
\begin{align}\label{TFDS}
\begin{split}
\ket{\psi(0)} :=  \frac{1}{\sqrt{Z_{\beta}}} \sum_{n} e^{-\frac{\beta E_n}{2}} \ket{n, n} \,, \qquad Z_{\beta}=\sum_n e^{-\beta E_n} \,,
\end{split}
\end{align}
where $Z_{\beta}$ is the partition function and $\beta = 1/T$ the inverse temperature.\footnote{The initial state implemented in the Householder reflections with the Hessenberg form is typically fixed as $(1,0,0,\cdots)^{T}$. As such, one should first perform a change of basis (i.e., ``rotation") from the desired initial vector to this one. For more details, we refer the readers to \cite{Balasubramanian:2022tpr}.}

Using the TFD state \eqref{TFDS} in certain chaotic systems, the authors in \cite{Balasubramanian:2022tpr} revealed the characteristic peak and plateau structure in spread complexity, which was conjectured to be a universal characteristic of chaotic systems. The similar structure has also been examined in both analytically and numerically in \cite{Erdmenger:2023wjg}. In this study, we intend to explore this characteristic by analyzing the spread complexity of ``integrable" quantum mechanics, especially of saddle-dominated scrambling systems.

\subsection{More on spread complexity: Ehrenfest theorem and early time growth}\label{}
\paragraph{Ehrenfest theorem.} 
An intriguing characteristic of the spread complexity is that the spread complexity can satisfy the Ehrenfest theorem~\cite{Erdmenger:2023wjg}: 
\begin{align}\label{EHRTH}
\begin{split}
\partial^2_t C(t) = - \left[ \left[ C(t)\,, \mathcal{L} \right], \mathcal{L} \right] \,,
\end{split}
\end{align}
where $\mathcal{L} = H \otimes \mathbb{I}$ is the Liouvillian, which is built with the identity $\mathbb{I}$. Furthermore, using \eqref{SEQ}-\eqref{SCDEF}, the Ehrenfest theorem \eqref{EHRTH} can also be expressed in terms of the Lanczos coefficients and transition amplitudes, i.e., 
\begin{align}\label{EHRTH2}
\begin{split}
\partial^2_t C(t) = 2 \sum_n  \left[ \left( b_{n+1}^2 - b_n^2 \right) \psi_n(t) \psi_n^{*}(t)  + \left(a_{n+1} - a_{n}\right)b_{n+1}  \psi_{(n+1}(t) \psi_{n)}^{*}(t)  \right]\,,
\end{split}
\end{align}
where $\mathcal{T}_{(a} \tilde{\mathcal{T}}_{b)}:=\frac{1}{2}\left(\mathcal{T}_{a}\tilde{\mathcal{T}}_{b}+\mathcal{T}_{b}\tilde{\mathcal{T}}_{a}\right)$.\footnote{In \eqref{EHRTH2}, we also corrected a sign error given in \cite{Erdmenger:2023wjg}.}
Note that \eqref{EHRTH2} is valid for any systems by construction.
In the main body of the manuscript, we will perform a numerical verification of the Ehrenfest theorem to substantiate this assertion.

\paragraph{The quadratic growth of spread complexity at early times.}
By solving the Schrödinger equation \eqref{SEQ} in the early-time regime, our objective here is to derive the analytical expression for spread complexity at early times, which can be used to validate our numerical findings in the following sections. A parallel analysis for the Krylov operator complexity can be found in \cite{Fan:2022xaa}.

To begin with, by imposing the boundary condition at $t=0$ in equation \eqref{SEQ}, one finds that 
\begin{align}\label{ANST}
\begin{split}
   \psi_{n}(0) &= \delta_{n0} \,, \\
 \dot{\psi}_{n}(0) &= -i \left(b_1 \delta_{n1} + a_0 \delta_{n0}\right) \,, \\
 \ddot{\psi}_{n}(0) &= - \left(a_0^2 + b_1^2\right) \delta_{n0} - a_0 b_1 \delta_{n1} - b_1 b_2 \delta_{n2} \,,
\end{split}
\end{align}
which yields
\begin{align}\label{ICSET}
\begin{split}
C(0) = 0 \,, \quad \dot{C}(0) = 0\,, \quad \ddot{C}(0) = 2 b_1^2 \,,
\end{split}
\end{align}
where $b_1^2$ originates from $|\dot{\psi}_{1}(0)|^2$ in \eqref{ANST}.

Then, the initial conditions \eqref{ICSET} together with the definition of spread complexity \eqref{SCDEF} lead to 
\begin{align}\label{ANAC}
\begin{split}
C(t) = b_1^2 t^2 + \mathcal{O}(t^3) \,.
\end{split}
\end{align}
Therefore, we find that the spread complexity shows \textit{quadratic} behavior in the early time regime. It is noteworthy that the Krylov operator complexity, as detailed in \cite{Fan:2022xaa}, also shares the same functional form as expressed in \eqref{ANAC}, i.e., the Lanczos coefficients $a_n$ does not play any role in early times.

\subsection{Example: quantum billiard system (chaotic vs. integrable system)}\label{}
Implementing the Lanczos algorithm in section \ref{SCLA}, we close this section with an illustrative (quantum mechanical) example of the spread complexity of \eqref{TFDS}, which can show the conjectured features -- the peak and plateau structure --  in chaotic systems~\cite{Balasubramanian:2022tpr}.

\begin{figure}[]
\centering
     \includegraphics[width=8.0cm]{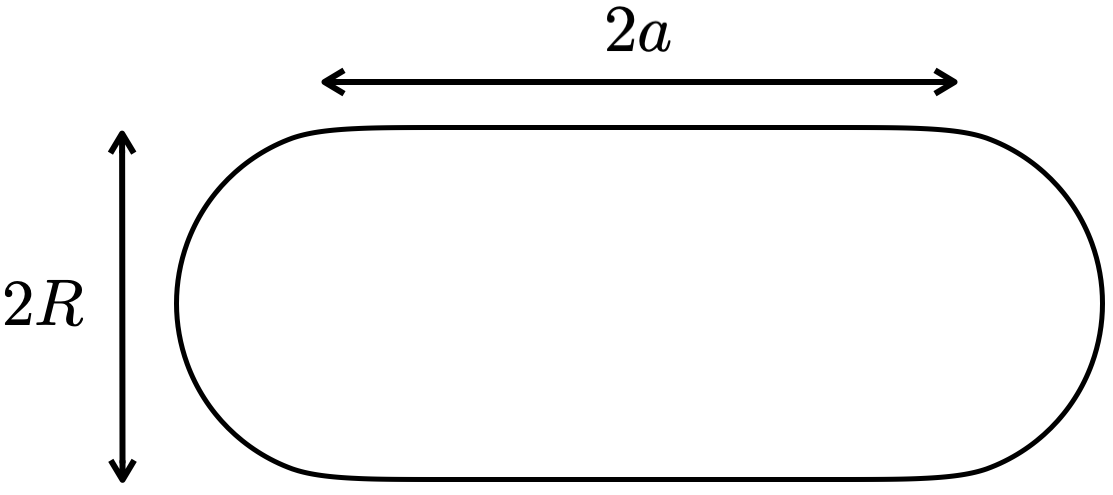}
 \caption{The domain of the billiard systems \eqref{BILLMO}.}\label{skaa}
\end{figure}

\paragraph{Quantum billiard systems.}
For this purpose, we consider the billiard systems~\cite{Sinai_1970,Bunimovich_1975,Bunimovich_1979,Bunimovich_1991,Benettin:1978aa} where its Hamiltonian is given as 
\begin{align}\label{BILLMO}
\begin{split}
  H = -\nabla^2  + V_{\text{billiard}}(x,y) \,,\qquad V_{\text{billiard}}(x,y) =  
   \begin{cases}
       0 & (x,y)\in \Omega \\
       \infty & \textrm{else}
   \end{cases}
\,,
\end{split}
\end{align}
where the domain $\Omega$ is depicted in Fig. \ref{skaa}.
Note that the billiard consists of semicircles with radii $R$ combined with straight lines of length $2a$.\footnote{This configuration yields its area as $A = \pi R^2 + 4 a R$. Here, we set $A=1$ in accordance with the existing literature, which extensively explores the study of Krylov and spread complexity~\cite{Hashimoto:2023swv,Camargo:2023eev} and the thermal OTOCs~\cite{Hashimoto:2017oit} within the same system.}

It is worth noting that the spread complexity of the billiard systems is studied in \cite{Hashimoto:2023swv} at infinite temperature ($\beta=0$). In what follows, we aim not only to demonstrate the consistency of our results with thier analysis when $\beta=0$, but also to explore the case of finite $\beta$.

Additionally, it is also pertinent to note that the parameter $a/R$ serves as a relevant deformation parameter in the context of billiards~\cite{Benettin:1978aa,McDonaldSpectrum,Casati1980} systems. Notably, the circle billiard ($a/R=0$) exhibits characteristics typical of an integrable system, such as a vanishing Lyapunov exponent~\cite{Hashimoto:2017oit}, while the stadium billiard ($a/R\neq0$) is regarded as non-integrable (i.e., chaotic systems), characterized by a finite Lyapunov exponent.

\paragraph{Spread complexity of billiard systems.}
Applying the Lanczos algorithm to the billiards systems \eqref{BILLMO}, we compute the spread complexity. Here, we set $a/R=1$ for the representative example for the stadium billiard.

\begin{figure}[]
 \centering
     \subfigure[The stadium billiard ($a/R=1$)]
     {\includegraphics[width=7.1cm]{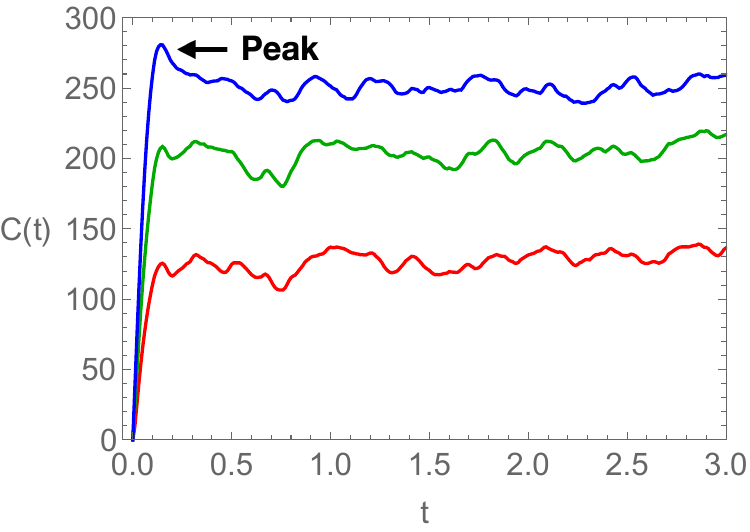} \label{}}
     \subfigure[The circle billiard ($a/R=0$)]
     {\includegraphics[width=7.1cm]{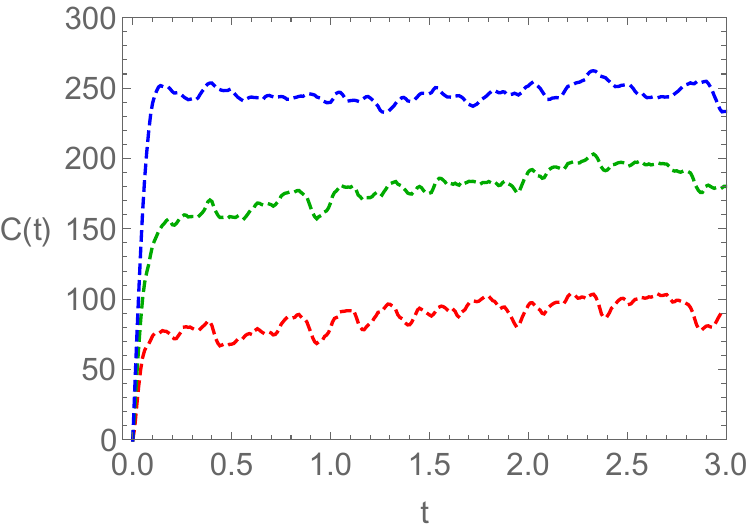} \label{}}
\caption{Spread complexity in billiard systems at $\beta = 0, 0.001, 0.003$ (blue, green, red). The left figure is for the chaotic system ($a/R=1$), while the right one is for the integrable system ($a/R=0$).}\label{SCQBS}
\end{figure}

In Fig. \ref{SCQBS}, we display the spread complexity for both $a/R=1$ and $a/R=0$.
Our main findings can be summarized in two aspects. 
Firstly, at $\beta=0$ (represented by the blue data), our results are consistent with those in \cite{Hashimoto:2023swv}, revealing the characteristic peak structure for the chaotic case ($a/R=1$).

Furthermore, we also find the effect of finite $\beta$ on spread complexity. We observe that a finite $\beta$ (from blue to red) not only suppresses the spread complexity in billiard systems, but also eliminates the characteristic peak. This finite $\beta$ effect aligns with observations in other chaotic systems, such as the Sachdev-Ye-Kitaev model or random matrix theory~\cite{Balasubramanian:2022tpr}.
\begin{figure}[]
 \centering
 \quad
     \subfigure[The stadium billiard ($a/R=1$)]
     {\includegraphics[width=6.5cm]{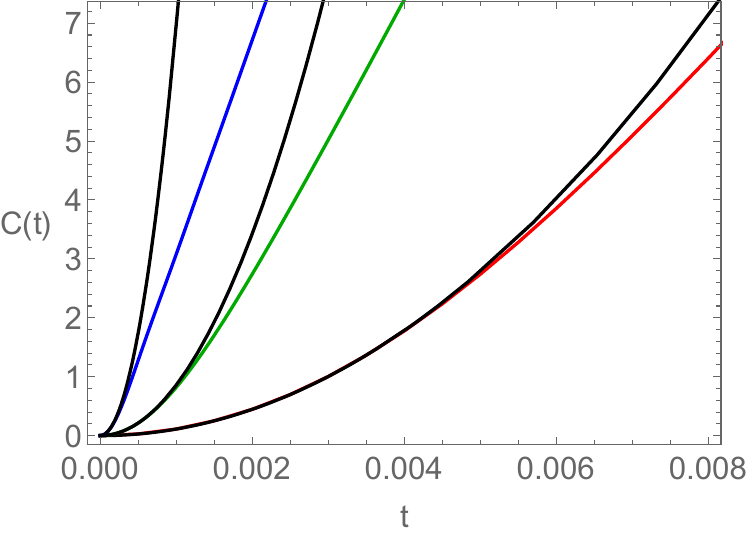} \label{}}
\quad
     \subfigure[The circle billiard ($a/R=0$)]
     {\includegraphics[width=6.5cm]{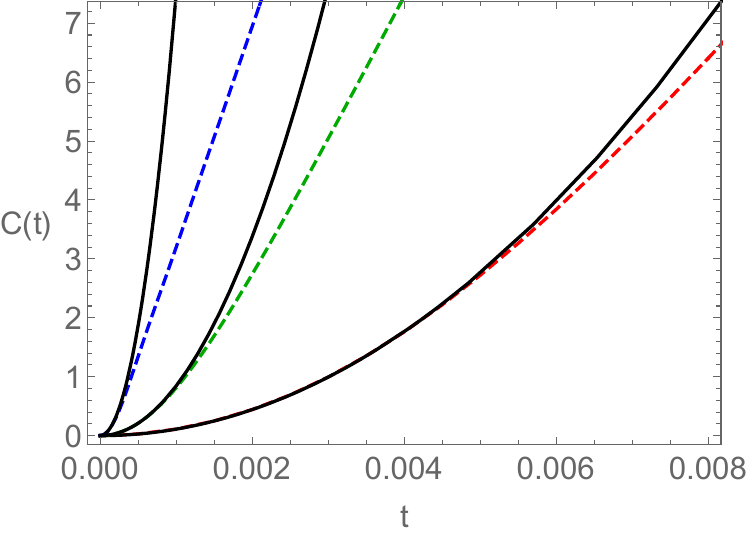} \label{}}
\caption{Spread complexity in billiard systems at early times. The numerical data (blue, green, red) corresponds to the data presented in Fig. \ref{SCQBS}. The black solid line is the analytic result \eqref{ANAC}.}\label{SCQBS2}
\end{figure}

In Fig. \ref{SCQBS2}, we also verify that our numerical results (blue, green, red) are in good agreement with the analytic expression (black), \eqref{ANAC}, at early times.\\

In the following sections, we investigate the spread complexity of non-chaotic systems characterized by saddle-dominated scrambling. Additionally, we explore the conjectured universal features of spread complexity for chaotic systems within this  class of systems.

\section{Spread complexity in the Lipkin-Meshkov-Glick model}\label{sec3}

In this section, we examine the spread complexity within an integrable model that features an unstable fixed point in its phase space, specifically focusing on the scrambling behavior around an unstable saddle point. The model under consideration here is a finite-dimensional quantum spin system known as the Lipkin-Meshkov-Glick (LMG) model~\cite{Lipkin:1964yk,GLICK1965211}.

\subsection{The LMG model}
We first review the LMG model in which in the classical limit, it is defined by the Hamiltonian as
\begin{align}\label{LMGH1}
\begin{split}
    H = x + 2 z^2 \,,
\end{split}
\end{align}
where the classical variables ($x\,,y\,,z$) obey the classical SU(2) algebra, expressed as $\{x\,,y\}=z$, subject to the constraint $x^2+y^2+z^2=1$. Here, $\{\,,\}$ denotes the Poisson bracket.

In order to find the saddle point, one can solve the Hamilton's equation of motion
\begin{align}
\begin{split}
   \frac{\dd X_i}{\dd t} = \{X_i \,, H\}\,,  \qquad  X_i := \{x\,,y\,,z\} \,.
\end{split}
\end{align}
For the Hamiltonian \eqref{LMGH1}, this equations can be expressed as
\begin{align}\label{HEOM}
\begin{split}
   \frac{\dd x}{\dd t} = -4 y z \,, \qquad \frac{\dd y}{\dd t} = -z + 4 x z \,, \qquad \frac{\dd z}{\dd t} = y \,,
\end{split}
\end{align}
where we used the following properties -- anti-commutativity and Leibniz's rule -- of the Poisson bracket
\begin{align}
\begin{split}
   \{f\,,g\} = -\{g\,,f\} \,, \qquad \{fg\,,h\} = \{f\,,h\}g + f\{g\,,h\} \,.
\end{split}
\end{align}

The saddle point is defined as the point ($x_s\,,y_s\,,z_s$) satisfying ${\dd X_i}/{\dd t} =0$. In other words, one can find the \textit{unstable} saddle point from \eqref{HEOM} as 
\begin{align}\label{USSP}
\begin{split}
   (x_s\,,y_s\,,z_s) = (1\,,0\,,0)\,.
\end{split}
\end{align}
Note that other saddle points can also be identified; nevertheless, based on the Jacobian of the transformation, it is straightforward to verify that these points are stable saddles.

It is also worth noting that around the unstable saddle point \eqref{USSP}
\begin{align}\label{}
\begin{split}
   (x\,,y\,,z) \,\,\rightarrow\,\, (1\,,0\,,0) + \epsilon \,(\delta x\,,\delta y\,,\delta z) \,,
\end{split}
\end{align}
the equations of motion \eqref{HEOM} can be linearized as
\begin{align}\label{}
\begin{split}
 \delta x'(t) = 0 \,, \qquad \delta y'(t) = 3 \delta z(t) \,, \qquad \delta z'(t) = \delta y(t) \,.
\end{split}
\end{align}
Subsequently, one can find the single equation 
\begin{align}\label{}
\begin{split}
 \delta y'(t) = \pm \lambda_L \, \delta y(t)  \quad\rightarrow\quad \delta y \,\approx\, e^{\pm \lambda_L t} \,, \quad \lambda_L = \sqrt{3} \,.
\end{split}
\end{align}
Therefore, the solution exhibits exponential growth characterized by a classical ``Lyapunov" exponent $\lambda_L$. However, it is imperative to note that this exponential behavior is \textit{not} indicative of chaos, as it only shows exponential growth in near the unstable saddle \eqref{USSP}.\\

The quantum mechanical version of the LMG model \eqref{LMGH1} can be given
\begin{align}\label{LMGH2}
\begin{split}
    H = \hat{x} + 2 \hat{z}^2 \,,
\end{split}
\end{align}
where $\{\hat{x}\,, \hat{y}\,, \hat{z}\} := \{\hat{S}_x/S\,, \hat{S}_y/S\,, \hat{S}_z/S\}$ are the rescaled SU(2) spin operators with spin $S$. They follow the commutation relations such as 
\begin{align}\label{}
\begin{split}
    [\hat{x}\,,\hat{y}] = i \hbar_{\text{eff}} \hat{z} \,,
\end{split}
\end{align}
where $\hbar_{\text{eff}} = 1/S$ is the effective Planck constant, which is giving the ``classical" limit ($\hbar_{\text{eff}}\rightarrow0$) at $S\rightarrow\infty$~\cite{Cotler:2017myn,Yin:2020oze}.

\subsection{Spread complexity and saddle-dominated scrambling}
Next, we compute the spread complexity within the LMG model \eqref{LMGH2}. The main objective is to examine the behavior and implications of saddle-dominated scrambling on the spread complexity.

Note that our analysis involves two free parameters: the spin $S$ and (inverse) temperature $\beta$. Similar to the prior investigation of Krylov operator complexity in the LMG model~\cite{Bhattacharjee:2022vlt}, we choose $S=25\,,50\,,$ and $75$ for spin. In addition, we primarily focus on the scenario of infinite temperature $\beta=0$ in the main text and discuss the finite $\beta$ effect at the end of the section.

\subsubsection{Lanczos coefficients}
\begin{figure}[]
 \centering
     \subfigure[$a_n$]
     {\includegraphics[width=7.1cm]{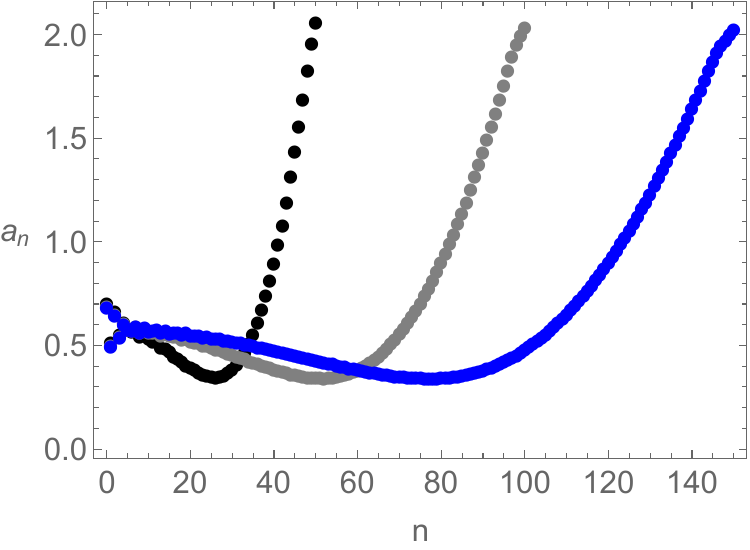} \label{}}
     \subfigure[$b_n$]
     {\includegraphics[width=7.1cm]{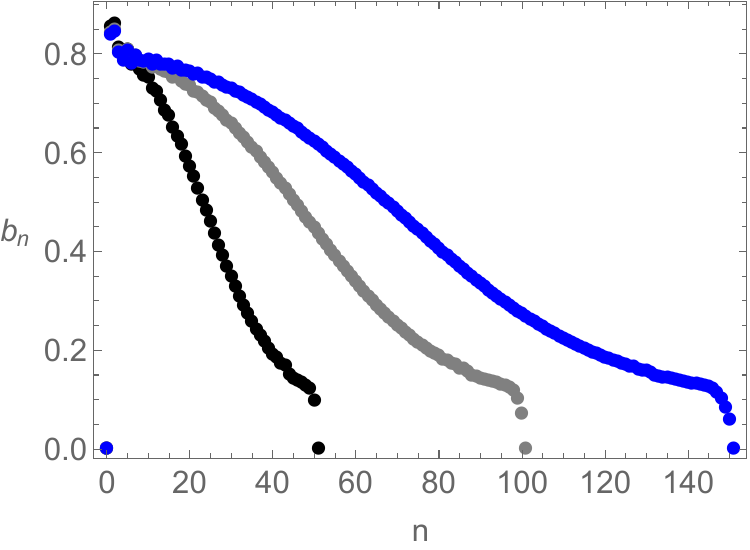} \label{}}
\caption{The Lanczos coefficients at $\beta=0$ with $S = {25,50,75}$ (black, gray, blue).}\label{LCLMGfig}
\end{figure}

Implementing the Lanczos algorithm given in the previous section, we present the Lanczos coefficients of LMG model in Fig. \ref{LCLMGfig}.
Here, we find two noteworthy observations. 

Initially, we find that $b_n$ exhibits analogous behavior to those observed in chaotic systems, wherein $b_n$ tends to vanish when $n$ approaches the size of the system $2S+1$. This suggests that, akin to the behavior identified in a chaotic model~\cite{Balasubramanian:2022tpr}, the Krylov basis can also span the full Hilbert space, even in the context of integrable systems when unstable saddles are included.

Regarding another Lanczos coefficients, $a_n$, we observe a non-monotonic behavior characterized by oscillations for small values of $n$. To the best of our knowledge, the behavior of $a_n$ is model-dependent, even for chaotic systems. For example, $a_n$ may exhibit oscillations near zero ($a_n\approx0$) in scenarios such as SYK models or Random Matrix Theory~\cite{Balasubramanian:2022tpr,Erdmenger:2023wjg}. It can also oscillate around finite values ($a_n\approx\#\gg1$) for systems like quantum billiards~\cite{Hashimoto:2023swv}. In addition, non-monotonic features are also reported in the chaotic limit of spin-1/2 models with disorder~\cite{Gautam:2023bcm}.

\subsubsection{Spread complexity}
Solving the Schrödinger equation \eqref{SEQ} together with the obtained Lanczos coefficients, we compute the spread complexity \eqref{SCDEF}.
\begin{figure}[]
\centering
{\includegraphics[width=7.1cm]{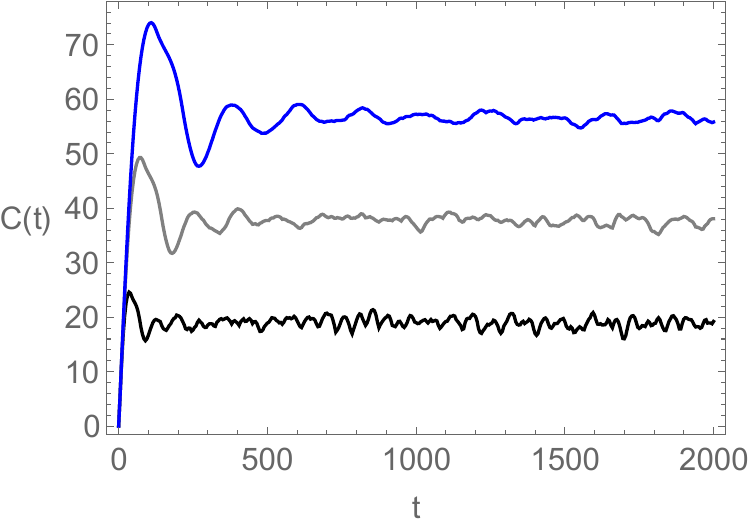}} 
\caption{The spread complexity at $\beta=0$ with $S = {25,50,75}$ (black, gray, blue).}\label{SCbeta0Fig} 
\end{figure}

\paragraph{The appearance of the peak in spread complexity.}
In Fig. \ref{SCbeta0Fig}, we present the time evolution of spread complexity. Notably, our findings indicate that the spread complexity of the LMG model can manifest conjectured features for chaotic systems: a ramp-``peak"-slope structure followed by a plateau~\cite{Balasubramanian:2022tpr}. This implies that spread complexity might lack the capacity to differentiate between saddle-dominated scrambling and generic chaos as did in OTOCs~\cite{Xu:2019lhc} and Krylov operator complexity~\cite{Bhattacharjee:2022vlt}.

Therefore, it is tempting to suggest that the proposed conjecture~\cite{Balasubramanian:2022tpr} would be revised to encompass not only quantum chaos but also the phenomena associated with saddle-dominated scrambling.\\

\paragraph{Spin $S$ dependence in spread complexity.}
We also explore another noteworthy feature of spread complexity within the LMG model. Especially, as depicted in Fig. \ref{SCbeta0Fig}, there can be a dependence on the spin variable $S$ observed in both the peak and saturation values.  We find that the fitting curves for the peak of the complexity $C_{\text{max}}$ and the saturation value $C(t=\infty)$ follow the expressions
\begin{align}\label{FCF}
\begin{split}
    C_{\text{max}} \approx S \,, \qquad C(t=\infty) \approx 3S/4 \,,
\end{split}
\end{align}
where we numerically evaluate $C(t)$ at $t=2000$ for $C(t=\infty)$.
\begin{figure}[]
 \centering
     \subfigure[$C_\text{max}$ vs. $S$]
     {\includegraphics[width=7.0cm]{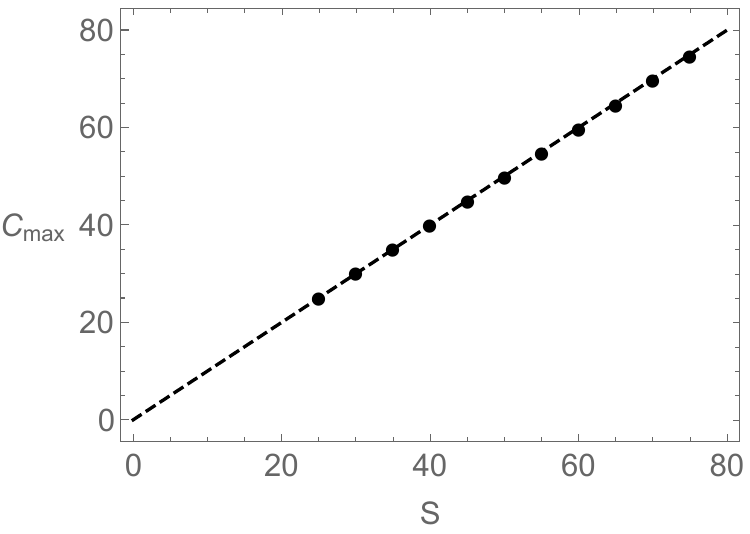} \label{}}
\quad
     \subfigure[$C(t=\infty)$ vs. $S$]
     {\includegraphics[width=7.3cm]{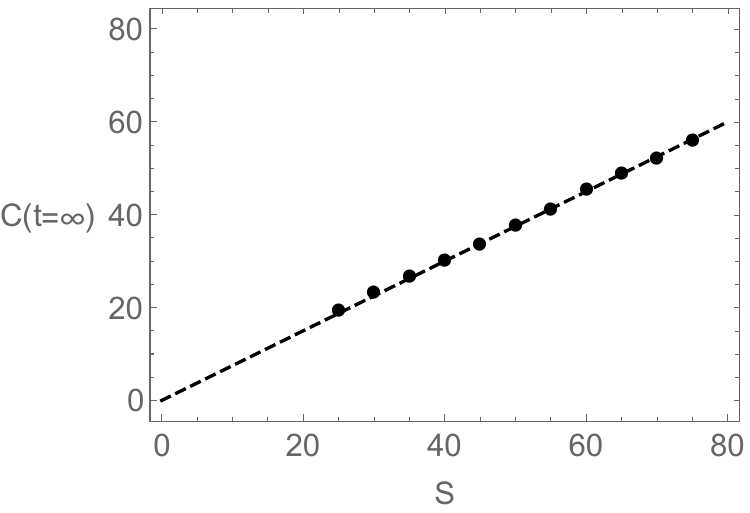} \label{}}
\caption{The spin $S$ dependence in spread complexity at $\beta=0$. The left panel represents the peak of complexity $C_\text{max}$ and the right panel is for the saturated value $C(t=\infty)$. The dashed lines are fitting curves \eqref{FCF}.}\label{FTC}
\end{figure}
See Fig. \ref{FTC}.\\

\begin{figure}[]
 \centering
      \subfigure[$S=75$]
     {\includegraphics[width=4.95cm]{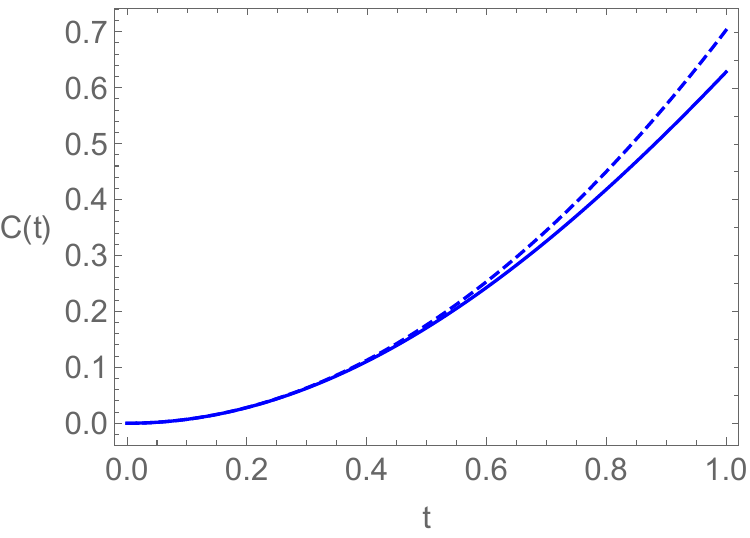}} 
      \subfigure[$S=50$]
     {\includegraphics[width=4.95cm]{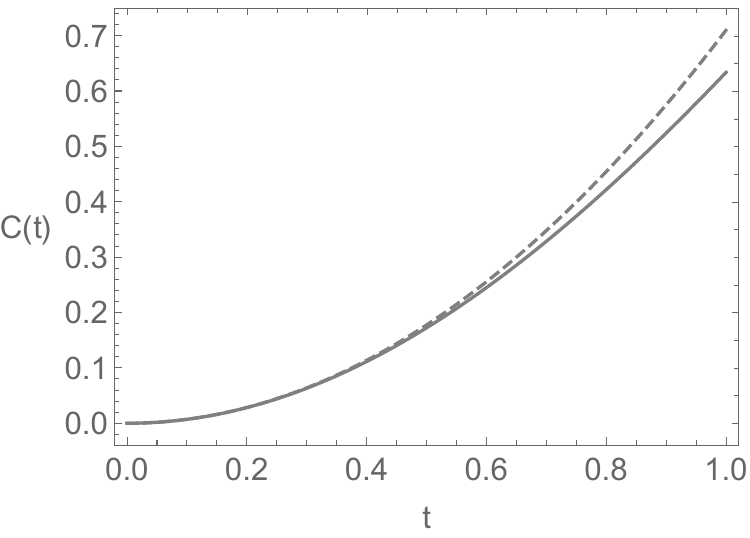}} 
      \subfigure[$S=25$]
     {\includegraphics[width=4.95cm]{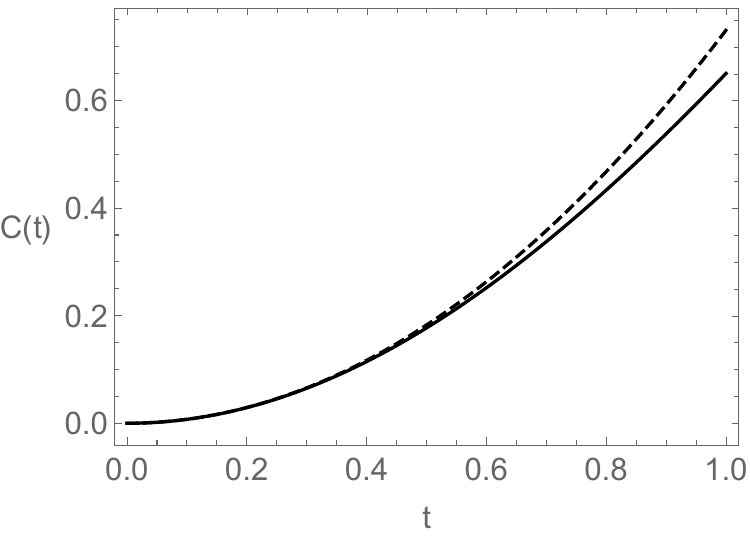}} 
     \caption{The spread complexity at $\beta=0$ with $S = {25,50,75}$ (black, gray, blue) in early-time regime. Solid lines are numerically evaluated data, while the dashed lines are analytic results \eqref{ANAC}.}\label{ETBC} 
\end{figure}

\paragraph{Early-time growth of complexity and Ehrenfest theorem.}
In addition to the late-time behavior of the spread complexity, $C(t=\infty)$, we also explore the opposite scenario: the early-time behavior. 
In Fig. \ref{ETBC}, we display the spread complexity in the early-time regime. 
The solid lines represent numerically evaluated data, while the dashed lines depict the analytic results \eqref{ANAC} derived in previous section \eqref{ANST}--\eqref{ANAC}.
We find that the spread complexity demonstrates quadratic early-time growth. Also, the same figure further highlights the excellent agreement between our numerical findings and analytic results.

In addition, we also validate the proposed Ehrenfest theorem \eqref{EHRTH2} in \cite{Erdmenger:2023wjg}, which establishes a relationship between the second-time derivative of spread complexity and a combination of Lanczos coefficients.
\begin{figure}[]
 \centering
      \subfigure[$S=75$]
     {\includegraphics[width=4.95cm]{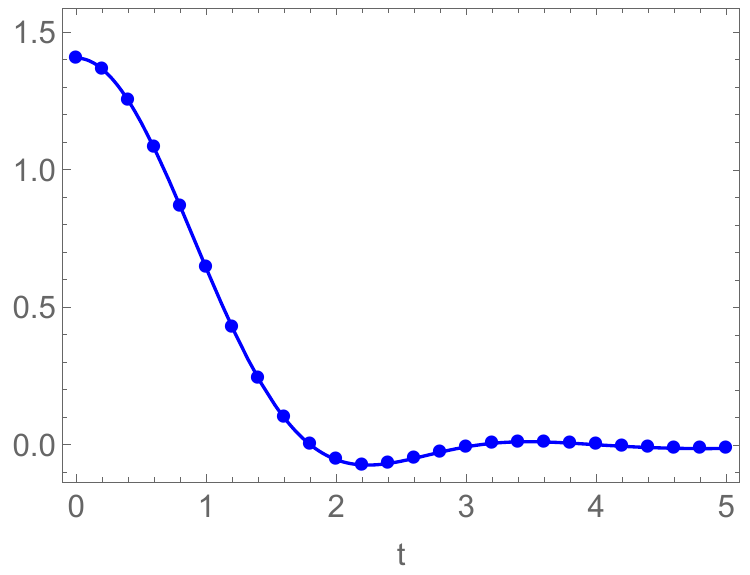}} 
      \subfigure[$S=50$]
     {\includegraphics[width=4.95cm]{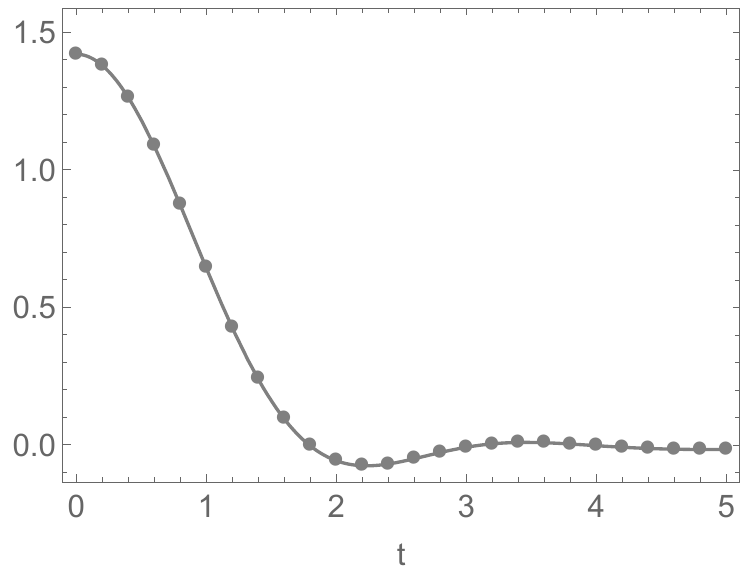}} 
      \subfigure[$S=25$]
     {\includegraphics[width=4.95cm]{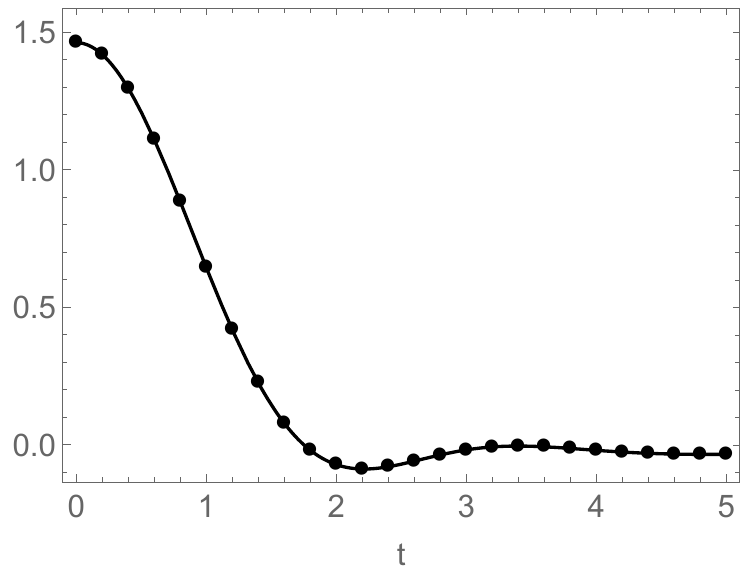}} 
     \caption{Ehrenfest theorem in the LMG model at $\beta=0$. The solid lines denote the L.H.S of \eqref{EHRTH2}, while the dots are the R.H.S of \eqref{EHRTH2}.}\label{EFTfig} 
\end{figure}
In  Fig. \ref{EFTfig}, the verification of the Ehrenfest theorem is demonstrated for the LMG model. This validation also serves to bolster the credibility of our numerical findings.\\

\paragraph{Spectral form factor and the transition probability.}
In addition to the conjectured universal behavior of spread complexity for Hamiltonians characterizing chaotic systems, the appearance of the peak~\cite{Balasubramanian:2022tpr}, there are two additional proposed arguments associated with the spread complexity of chaotic systems. 
The first pertains to insights derived from the \textit{Spectral Form Factor} (SFF)~\cite{Balasubramanian:2022tpr,Erdmenger:2023wjg}, while the second is related to observations based on the \textit{transition probability}~\cite{Erdmenger:2023wjg}.

For the case of SFF, when the system is chaotic, it is argued \cite{Erdmenger:2023wjg} that the quadratic growth, linear growth, a peak, and saturation of complexity are in analogy to the slope, dip, ramp and plateau of the spectral form factor. This analogy is established by considering the close proximity of the transition time scales in both phenomena. 

\begin{figure}[]
 \centering
     \subfigure[Spread complexity]
     {\includegraphics[width=7.0cm]{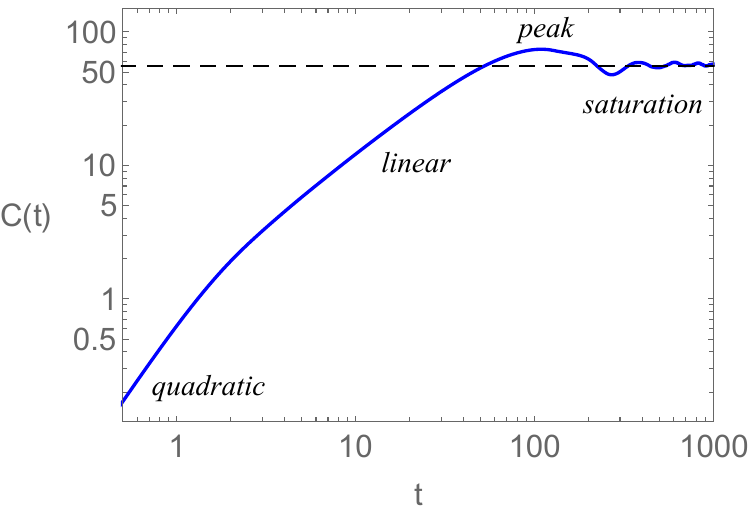}\label{LMGComvsSFFa}}
\qquad
     \subfigure[Spectral Form Factor (SFF)]
     {\includegraphics[width=7.1cm]{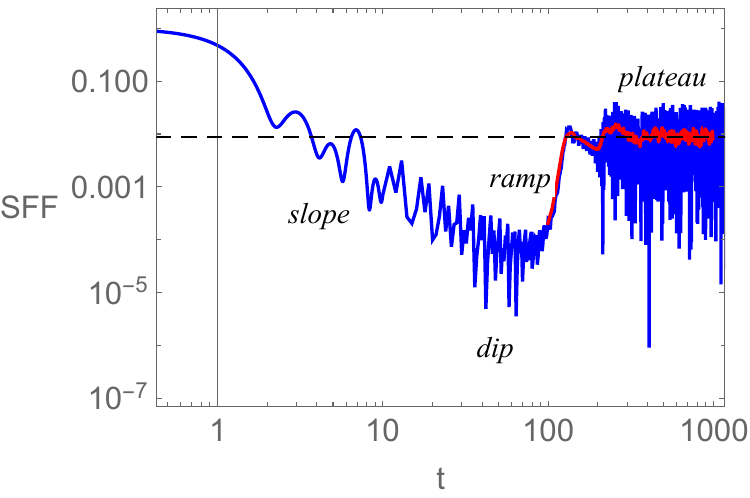} \label{LMGComvsSFFb}}
\caption{Spread complexity and Spectral Form Factor in the LMG model when $\beta=0$ and $S=75$. \textbf{Left:} the spread complexity. The dashed line is $C(t=\infty)$ in \eqref{FCF}. \textbf{Right:} Spectral Form Factor. The dashed line is determined by \eqref{PROLARGE} where $ \text{SFF} = \left( |\psi_0|^2 \right)_{t=\infty}$.  The red data is the average of the noisy signal in the ramp and plateau, obtained by \textit{Mathematica} built-in function: MovingAverage. }\label{LMGComvsSFF}
\end{figure}

In Fig. \ref{LMGComvsSFF}, we find that the LMG model can indeed demonstrate the proposed resemblance between spread complexity and the SFF. In particular, our result suggests that the linear growth in spread complexity might be determined by the slope of SFF, which is not be exclusively tied to chaotic behavior. This observation aligns with the findings reported in \cite{Erdmenger:2023wjg}.
This suggests that the analogical relationship between spread complexity and SFF may also manifest in integrable systems featuring a saddle point.\\

In addition to the connection between the linear growth and the slope, one can find the further relationship between the spread complexity and SFF using the transition probability $|\psi_n|^2$ in \eqref{SCDEF}. This probability is given by the wave function, specifically the norm of the amplitude squared, within the Krylov space.

\begin{figure}[]
 \centering
      \subfigure[$n=0$]
     {\includegraphics[width=4.95cm]{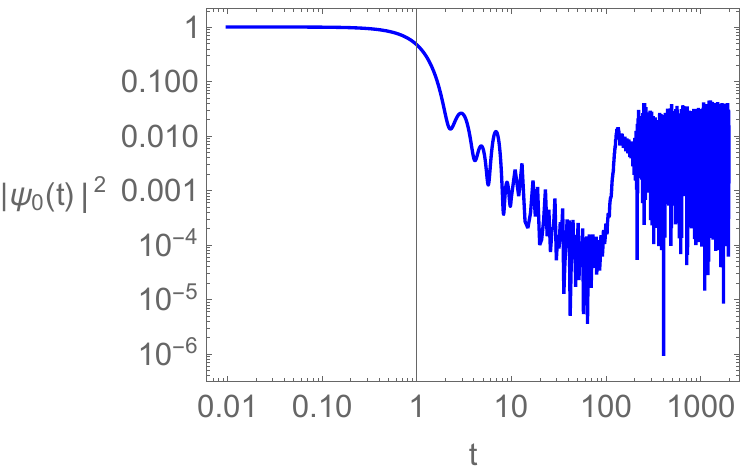}\label{LMGprobfiga}}
      \subfigure[$n=5$]
     {\includegraphics[width=4.95cm]{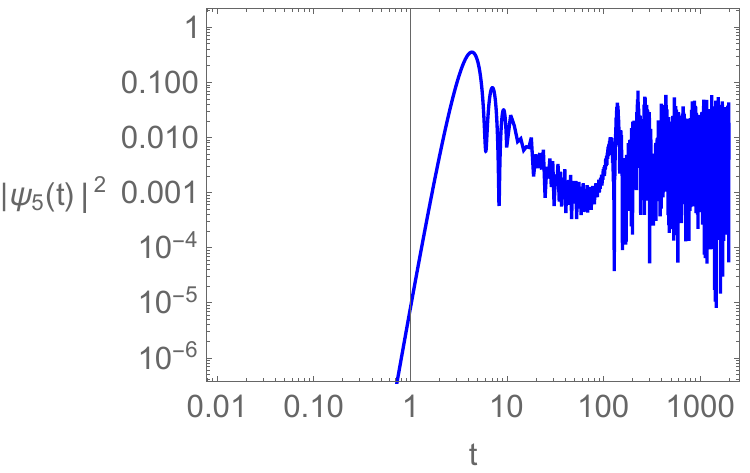}\label{}} 
      \subfigure[$n=10$]
     {\includegraphics[width=4.95cm]{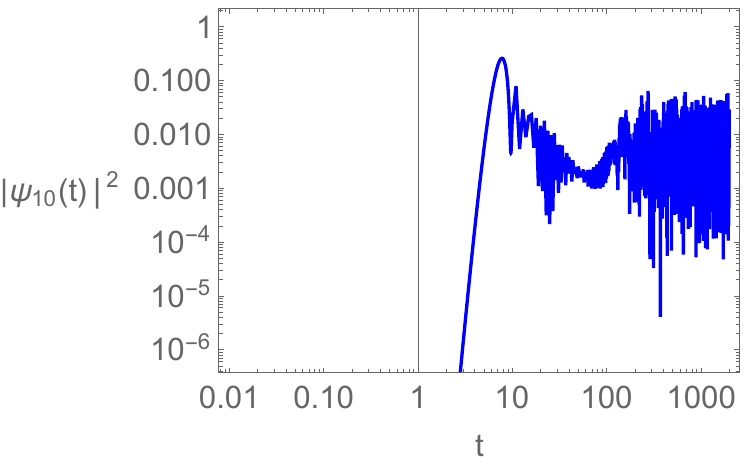}\label{}} 
     \caption{The transition probability $|\psi_n|^2$ of the LMG model when $\beta=0$ and $S=75$.}\label{LMGprobfig} 
\end{figure}

To begin with, it has been demonstrated \cite{Erdmenger:2023wjg} that, in the context of the spread complexity of the TFD state, $|\psi_0|^2$ corresponds to the SFF.
In order to illustrate this relationship for the LMG model, we present $|\psi_n|^2$ for various value of $n$ in Fig. \ref{LMGprobfig}: one can find that the case with $n=0$, Fig. \ref{LMGprobfiga}, is equivalent to the SFF in Fig. \ref{LMGComvsSFFb}.

By considering $|\psi_0|^2$ or equivalently SFF, it has been argued~\cite{Erdmenger:2023wjg} that the presence of a ``long" (or ``clear") ramp in the SFF may be responsible to the occurrence of a peak in spread complexity after its linear growth. Our observation of a distinct ramp in Fig. \ref{LMGComvsSFFb} supports the notion that this argument may hold true even for integrable systems with saddles.

Furthermore, in addition to the presence of a peak in spread complexity, the saturated value can also be associated with the plateau in the SFF. This connection is established through the following relationship.

For the case of TFD state at $\beta=0$, it has been demonstrated~\cite{Erdmenger:2023wjg,Rabinovici:2020ryf,Rabinovici:2022beu} that, in the late-time regime, the transition probability and the spread complexity can be expressed to obey 
\begin{align}\label{PROLARGE}
\begin{split}
    \left( |\psi_0|^2 \right)_{t=\infty} = \frac{1}{1 + 2\,C(t=\infty)}  \,.
\end{split}
\end{align}
These results hold independently of the presence of chaotic behavior and are essentially a consequence of considering the maximally entangled state as the reference state~\cite{Erdmenger:2023wjg}.
We numerically validate this relation \eqref{PROLARGE} in the LMG model, as depicted in dashed lines in Fig. \ref{LMGComvsSFF}.\\

\begin{figure}[]
 \centering
     {\includegraphics[width=7.1cm]{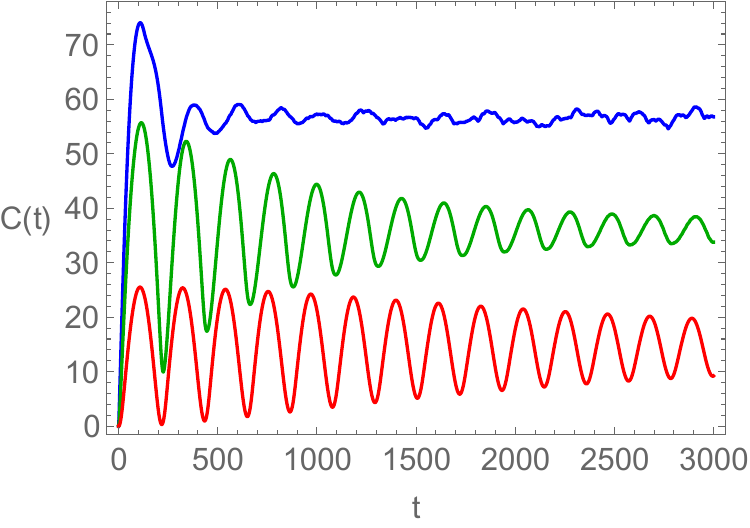}} 
\caption{The spread complexity at $S=75$ when $\beta = {0\,,3\,,10}$ (blue, green, red).} \label{LMGfinitebetaFig}
\end{figure}

\paragraph{Effect of finite temperature.}
Last but not least, we also address the finite temperature $\beta$ effect on the spread complexity in the LMG model. Similar to observations in previous investigations of spread complexity in chaotic models (e.g., SYK or random matrix theory~\cite{Balasubramanian:2022tpr}, or the quantum billiard problems discussed in the previous section), we note that a finite $\beta$ not only suppresses the spread complexity but also eradicates the characteristic peak. This is illustrated in Fig. \ref{LMGfinitebetaFig}.
%

{
It is worth noting that the diminishment of the characteristic feature of quantum choas at lower temperatures (i.e., finite $\beta$) may not be a surprise. Analogous phenomena have been observed in related studies, such as the transition of the exponential growth of out-of-time-ordered correlators (OTOCs) to oscillatory behavior with decreasing temperature: e.g., see Fig. 13 in \cite{Akutagawa:2020qbj}.
}

\section{Spread complexity in the inverted harmonic oscillator}\label{sec4}
In this section, we study the spread complexity within another toy model for saddle-dominated scrambling: the inverted harmonic oscillator.\footnote{It is noteworthy that the inverted harmonic oscillator may hold potential physical significance. For instance, a relativistic particle near a black hole horizon (being pulled by a force towards the exterior) may experience an inverted harmonic oscillator potential, contributing to chaotic behavior near the horizon~\cite{Hashimoto:2016dfz}.} Specifically, our focus is on one-dimensional quantum mechanics with a potential whose part is an inverted harmonic oscillator. Within the same model, OTOC has been explored in \cite{Hashimoto:2020xfr,Bhattacharyya:2020art}, while Krylov operator complexity in a related context has been studied in \cite{Baek:2022pkt}.

\subsection{The inverted harmonic oscillator model}
\begin{figure}[]
 \centering
     {\includegraphics[width=7.1cm]{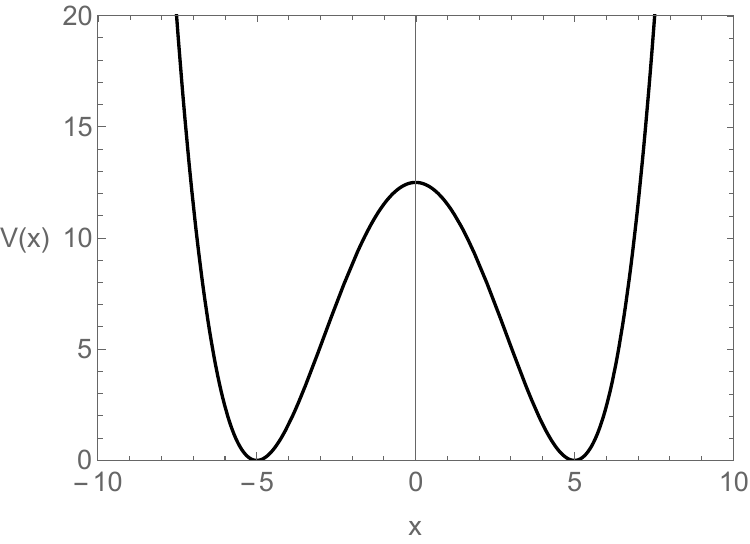}} 
\caption{Potential $V(x)$ of inverted harmonic oscillator models \eqref{IHOpotential} when $\lambda = 2,\, g=1/50$. }\label{IHOG}
\end{figure}

Let us first review the quantum mechanical system including an inverted harmonic oscillator~\cite{Hashimoto:2020xfr}. The system is given by the Hamiltonian:
\begin{align}\label{IHOpotential}
\begin{split}
H &:= p^2 + V(x)\,, \qquad V(x) := g\left(x^2 - \frac{\lambda^2}{8g}\right)^2 =  -\frac{1}{4} \lambda^2 x^2 + g x^4 + \frac{\lambda^4}{64g}\,.
\end{split}
\end{align}
Here $\lambda$ and $g$ serve as constant parameters that define the potential's shape. Note that the potential $V(x)$ corresponds to the Higgs potential in the high-energy theoretical context. In this paper, following the previous studies~\cite{Hashimoto:2020xfr,Baek:2022pkt}, we choose $\lambda = 2,\, g=1/50$ and consider the truncation (or maximum) eigenstates at $i = 200$: $H \ket{i} = E_i \ket{i}$. The corresponding potential is visualized in Fig. \ref{IHOG}.

Two remarks are in order.
First, although this model is classically non-chaotic (or regular), instability can arise around $x=0$, the unstable maximum of the potential. Consequently, a non-vanishing ``Lyapunov exponent" exists in this region, equal to the parameter $\lambda$ in the potential \eqref{IHOpotential}. Also note that $\lambda$ determines the curvature of the unstable peak.\\
Second, the inclusion of the $x^4$ term is essential for determining a well-defined energy~\cite{Hashimoto:2020xfr,Ali:2019zcj,Hashimoto:2017oit}. For instance, through an analytic continuation of the frequency $\omega$, the standard harmonic oscillator can be transformed into the inverted harmonic oscillator. However, such a naive analytic continuation results in a purely imaginary energy (and introduces ambiguity in defining thermal OTOCs~\cite{Hashimoto:2020xfr}). To address this and ensure proper definitions, one needs to consider cases with bounded potentials by introducing the $x^4$ term.

\subsection{Spread complexity in the inverted harmonic oscillator}
Here, we provide a summary of the numerical findings of spread complexity for the inverted harmonic oscillator model \eqref{IHOpotential}. Essentially, the results are similar to those obtained for the LMG model discussed in the previous section.

\begin{figure}[]
 \centering
     \subfigure[$a_n$]
     {\includegraphics[width=7.1cm]{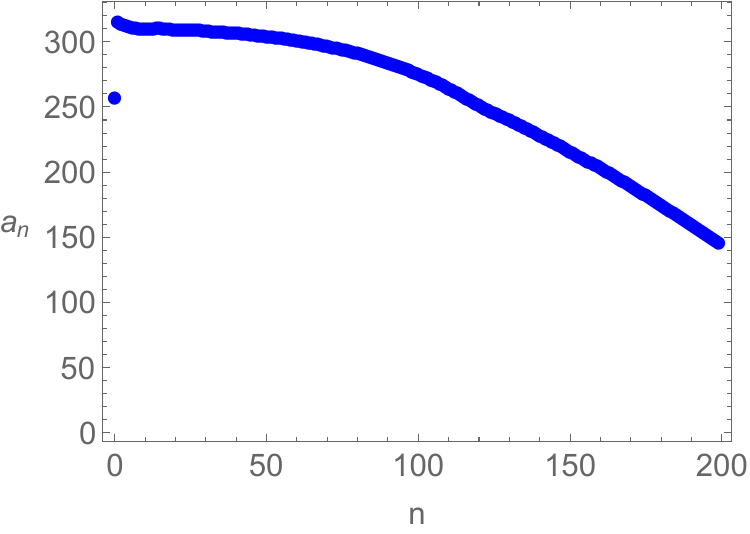} \label{}}
     \subfigure[$b_n$]
     {\includegraphics[width=7.1cm]{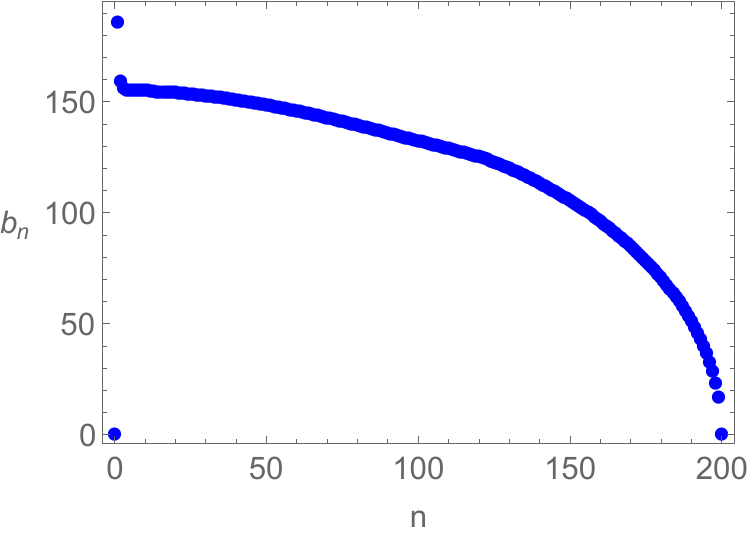} \label{}}
\caption{The Lanczos coefficients of the inverted harmonic oscillator model at $\beta=0$.}\label{LanczosIHOfig}
\end{figure}

The Lanczos coefficients are presented in Fig. \ref{LanczosIHOfig}.
Similar to the LMG model, the second kind of Lanczos coefficients ($b_n$) tends to vanish as $n$ approaches the system size. In contrast, the first kind ($a_n$) exhibits a more monotonic behavior compared to the LMG model.

\begin{figure}[]
\centering
{\includegraphics[width=7.1cm]{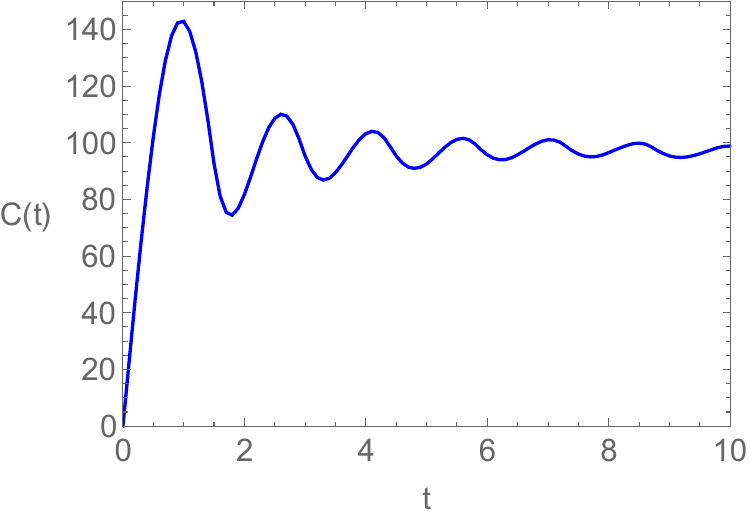}} 
\caption{The spread complexity of the inverted harmonic oscillator model at $\beta=0$.}\label{SCOMIHO} 
\end{figure}

We also display the spread complexity of the inverted harmonic oscillator in Fig. \ref{SCOMIHO}, showcasing characteristics conjectured for chaotic systems, specifically the emergence of a peak. 
Consequently, our examination of the inverted harmonic oscillator model serves as an additional example suggesting a potential refinement of the proposed conjecture~\cite{Balasubramanian:2022tpr}. This refinement would extend the conjecture to encompass not only quantum chaos but also phenomena associated with saddle-dominated scrambling.

\begin{figure}[]
 \centering
      \subfigure[Early-time behavior of spread complexity]
     {\includegraphics[width=7.1cm]{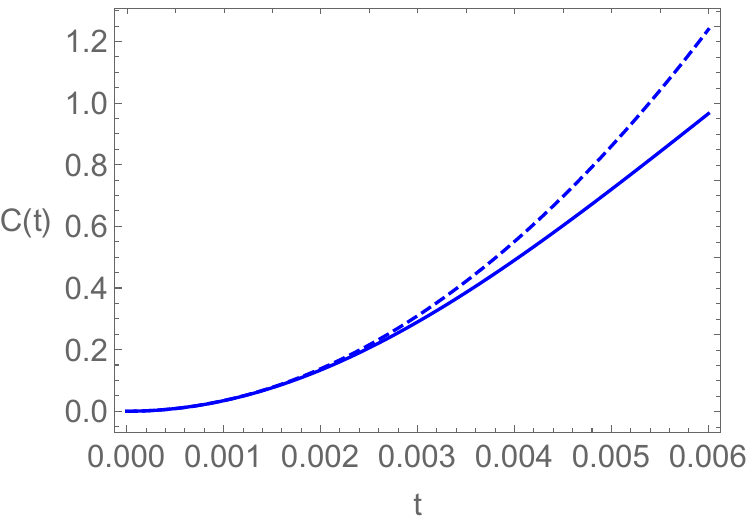}} 
\quad
      \subfigure[Verification of the Ehrenfest theorem]
     {\includegraphics[width=7.1cm]{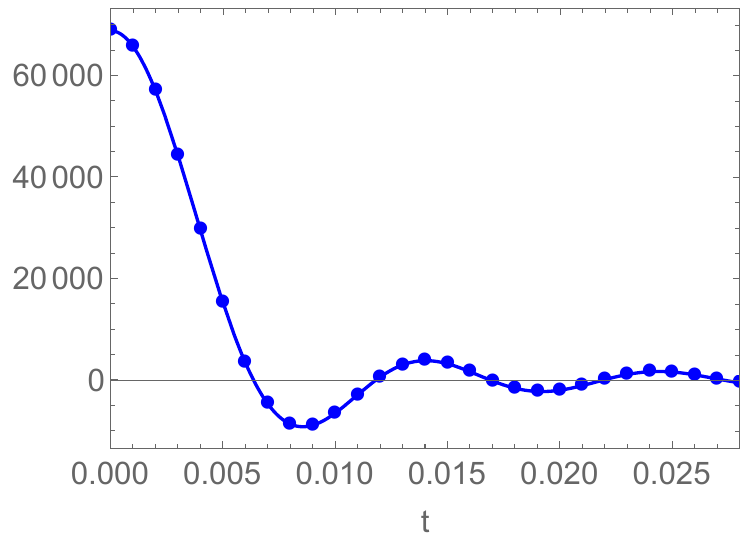}} 
     \caption{\textbf{Left:} The spread complexity of the inverted harmonic oscillator model at $\beta=0$ in early-time regime. Solid line is a numerical result and the dashed line is an analytic result \eqref{ANAC}. \textbf{Right:} Ehrenfest theorem in the inverted harmonic oscillator model at $\beta=0$. The solid lines denote the L.H.S of \eqref{EHRTH2}, while the dots are the R.H.S. of \eqref{EHRTH2}.}\label{IHOanavsnum} 
\end{figure}

Next, in Fig. \ref{IHOanavsnum}, we validate our numerical results of spread complexity with analytical expressions. The left panel in Fig. \ref{IHOanavsnum} focuses on the early-time behavior of complexity \eqref{ANAC}, while the right panel pertains to the Ehrenfest theorem \eqref{EHRTH2}.

\begin{figure}[]
 \centering
      \subfigure[Spread complexity]
     {\includegraphics[width=4.75cm]{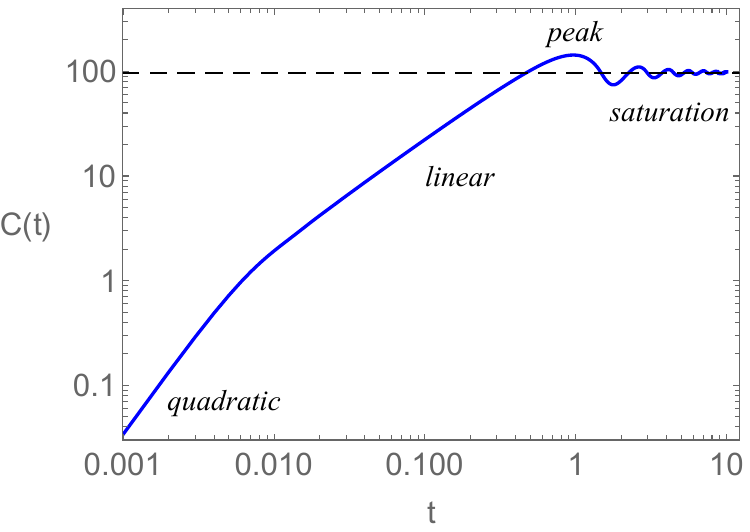}\label{}}
      \subfigure[Spectral Form Factor]
     {\includegraphics[width=4.85cm]{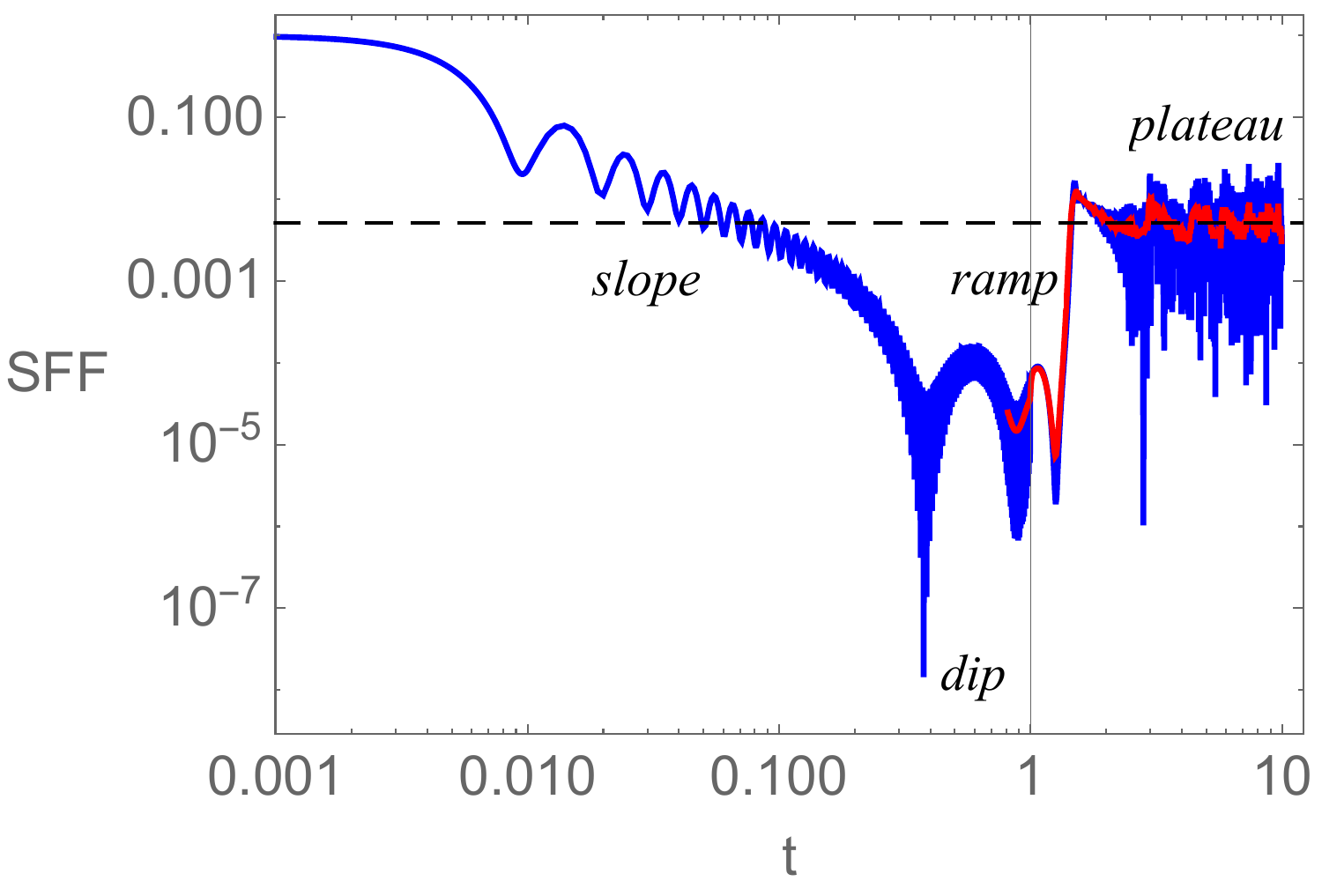}\label{}} 
      \subfigure[Transition probability]
     {\includegraphics[width=5.15cm]{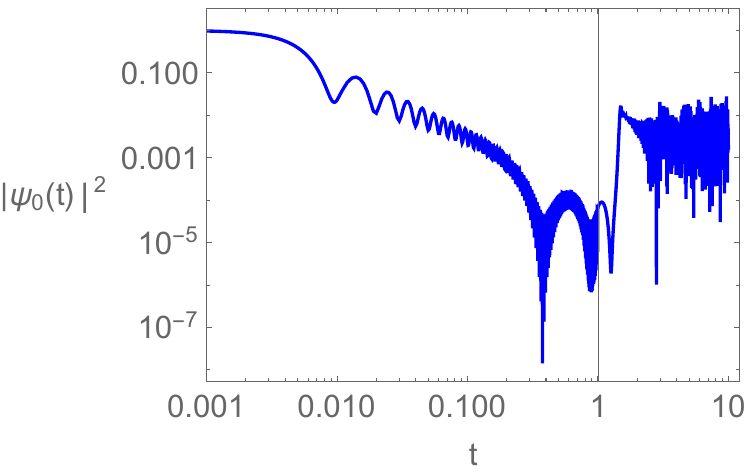}\label{}} 
     \caption{Spread complexity, Spectral Form Factor (SFF), and transition probability of the inverted harmonic oscillator model at $\beta=0$. The dashed lines are \eqref{PROLARGE}. The red data is the average of the noisy signal in the ramp and plateau.}\label{remainedplots} 
\end{figure}

Considering the Spectral Form Factor (SFF), we find its relationship, as suggested in \cite{Erdmenger:2023wjg}, with spread complexity. The identified connections include:
(I) The transition time scale in spread complexity closely aligns with that in the SFF;
(II) SFF is shown to be equivalent to the transition probability $|\psi_0|^2$;
(III) The presence of a distinct ramp in the SFF may be a contributing factor to the occurrence of a peak in spread complexity;
(IV) The late-time behavior follows the expression \eqref{PROLARGE}.\\
These observations are depicted in Fig. \ref{remainedplots}.

\begin{figure}[]
 \centering
     {\includegraphics[width=7.1cm]{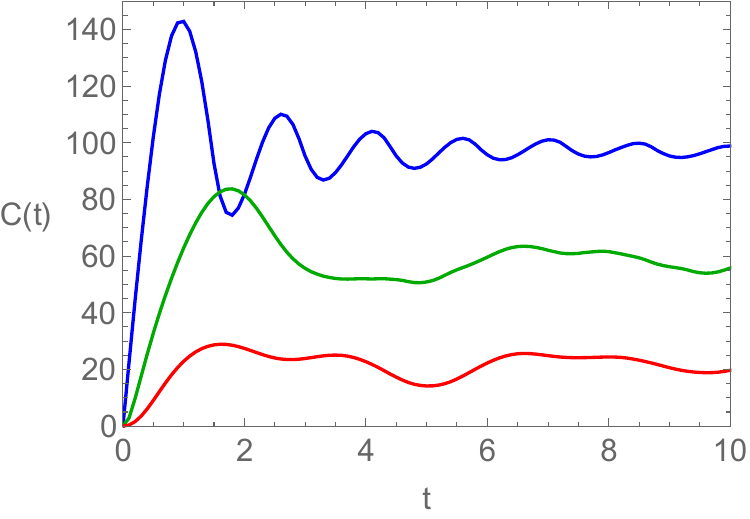}} 
\caption{The spread complexity of the inverted harmonic oscillator model when $\beta = {0\,,0.05\,,0.15}$ (blue, green, red).} \label{IHObetafig}
\end{figure}

Finally, we illustrate the temperature ($\beta$) dependence in spread complexity in Fig. \ref{IHObetafig}. 
The results indicate that with an increase in $\beta$:
(I) the spread complexity is suppressed;
(II) the characteristic peak diminishes.

\section{Conclusions and outlook}\label{sec5}
We investigated the spread complexity of the thermofield double state within integrable systems featuring saddle-dominated scrambling. Our focus was on two representative toy models: the LMG model and the inverted harmonic oscillator. The main findings of our study are summarized as follows.

Utilizing the Lanczos algorithm, our numerical analyses revealed that the spread complexity of our toy models exhibits conjectured features typical of chaotic systems, characterized by a ramp-peak-slope-plateau structure~\cite{Balasubramanian:2022tpr}. This observation suggests that, in the presence of unstable saddle points, spread complexity may face challenges in distinguishing between saddle-dominated scrambling and genuine quantum chaos, similar to other measures such as OTOC~\cite{Xu:2019lhc} and Krylov operator complexity~\cite{Bhattacharjee:2022vlt}. In essence, our results imply that the proposed conjecture of spread complexity~\cite{Balasubramanian:2022tpr} may benefit from refinement or additional physical input to unequivocally diagnose the presence or absence of chaos.

Additionally, we explored other aspects of spread complexity, focusing on two proposals related to chaotic systems. The first involves insights from the spectral form factor (SFF) \cite{Balasubramanian:2022tpr, Erdmenger:2023wjg}, while the second is associated with observations based on transition probability \cite{Erdmenger:2023wjg}.
Our results suggest that the analogous relationship between spread complexity and SFF—where the transition time scales in both quantities are similar—may also manifest in integrable systems featuring a saddle point. Moreover, we discovered that the late-time behavior of spread complexity is directly linked to that of the SFF \eqref{PROLARGE}, where the latter is equivalent to the transition probability within the Krylov space. Our findings align with the argument in~\cite{Erdmenger:2023wjg}, indicating that a clear ramp in the transition probability can lead to a peak in spread complexity, even in integrable systems with unstable saddle points

In addition, we validated our numerical findings through a comparison with analytic results for spread complexity. Firstly, we numerically confirmed the Ehrenfest theorem for complexity~\cite{Erdmenger:2023wjg}. Furthermore, by solving the Schrödinger equation in the early-time regime, we derived the analytic expression for spread complexity \eqref{ANAC}. Our verification indicated that spread complexity exhibits quadratic behavior in the early-time regime, with its growth rate corresponding to one of the Lanczos coefficients. The consistency between our numerical results and the analytic quadratic growth further strengthens the reliability of our findings.

There are several interesting ideas to explore in the context of this work. First of all, the phenomenon of saddle-dominated scrambling extends beyond the confines of the LMG model or the inverted harmonic oscillator. Consideration can be given to diverse models demonstrating such behavior, exemplified by the quantum Dicke model~\cite{Dicke:1954zz}, where significant research has been dedicated to the investigation of scrambling and OTOCs~\cite{Lewis-Swan:2018sdr,Chavez-Carlos:2018ijc,Wang:2018tmi}. Another illustrative instance is the Feingold-Peres (FP) model~\cite{Feingold:1983bw,Feingold:1984aa,Peres:1984zz}, which demonstrates saddle-dominated scrambling despite being classically chaotic. Therefore, by further exploration of spread complexity within these models, one could not only confirm whether the observed peak is a generic feature or limited to specific model classes, but also gain a more complete understanding of the role that spread complexity plays in situations involving saddle-dominated scrambling.

It is also instructive to recall that, in the historical context, the exploration of chaos and complexity has been a focal point within the framework of the AdS/CFT correspondence and black hole physics. In particular, several conjectures posit that the complexity of the holographic quantum system can be a useful probe of the interior of the dual black hole~\cite{Stanford:2014jda,Brown:2015bva,Brown:2015lvg,Couch:2016exn,Belin:2021bga} and may explain the emergence of spacetime \cite{Czech:2017ryf,Susskind:2019ddc,Pedraza:2021mkh,Pedraza:2021fgp,Pedraza:2022dqi,Carrasco:2023fcj}.
One outstanding issue in this context is the ambiguity arising from multiple proposals defining complexity. In contrast, Krylov (operator/state) complexity provides an unambiguous definition. A crucial question emerges: can Krylov complexity accurately describe the growth of a black hole interior? Related studies in this direction can be found in~\cite{Kar:2021nbm}.
Within this perspective, understanding the relationship between conventional computational complexity and Krylov complexity becomes a significant concern. While proposals exist for the gravity dual interpretation of conventional circuit complexity, recent investigations have explored a dual interpretation of Krylov complexity~\cite{Rabinovici:2023yex}. A more comprehensive examination of Krylov complexity across diverse quantum mechanical scenarios may shed light on the enigma of black hole interiors and the emergence of spacetime.

We leave all these interesting subjects for future work and hope to come back to them in the near future.

\acknowledgments
We would like to thank {Hugo A. Camargo, Johanna Erdmenger, Viktor Jahnke, Shao-Kai Jian, Keun-Young Kim, Mitsuhiro Nishida, and Zhuo-Yu Xian} for valuable discussions and correspondence. 
KBH is supported by the Basic Science Research Program
through the National Research Foundation of Korea (NRF) funded by the Ministry of Science, ICT $\&$ Future Planning (Grant No.NRF-2021R1A2C1006791) and GIST Research Institute
(GRI) grant funded by the GIST in 2023.
HSJ and JFP are supported by the Spanish MINECO ‘Centro de Excelencia Severo Ochoa' program under grant SEV-2012-0249, the Comunidad de Madrid ‘Atracci\'on de Talento’ program (ATCAM) grant 2020-T1/TIC-20495, the Spanish Research Agency via grants CEX2020-001007-S and PID2021-123017NB-I00, funded by MCIN/AEI/10.13039/501100011033, and ERDF A way of making Europe.
%


\bibliographystyle{JHEP}

\providecommand{\href}[2]{#2}\begingroup\raggedright\endgroup

\end{document}